\documentstyle[12pt]{article}

\def \be{\begin{equation}}
\def \o{\over}
\def \ee{\end{equation}}
\def \eea{\end{eqnarray}}
\def \bea{\begin{eqnarray}}
\def \l{\label}
\begin{document}
\begin{titlepage}
\vspace*{5mm}
\begin{center}{\Large \bf Massive Schwinger model and its confining \\
\vskip 1cm
aspects on curved space--time }
\vskip 1cm
{\bf M. Alimohammadi$^{a,b}$ \footnote {e-mail:alimohmd@theory.ipm.ac.ir},
H. Mohseni Sadjadi$^a$ \footnote {e-mail:amohseni@khayam.ut.ac.ir}}
\vskip 1cm
{\it $^a$Physics Department, University of Tehran, North Karegar,} \\
{\it Tehran, Iran }\\
{\it $^b$Institute for Studies in Theoretical Physics and Mathematics,}\\
{\it  P.O.Box 5531, Tehran 19395, Iran}

\end{center}
\vskip 2cm

\begin{abstract}
Using a covariant method to regularize the composite
operators, we obtain the bosonized action of the massive Schwinger
model on a classical curved background. Using the solution of the
bosonic effective action, the energy of two static external
charges with finite and large distance separation on a static
curved space--time is obtained. The confining behavior of this
model is also explicitly discussed.

\end{abstract}
\end{titlepage}
\newpage

\section{Introduction}

Two dimensional quantum electrodynamics or the Schwinger model \cite{sc} may
be served as a laboratory for studying four dimensional gauge theories
and important phenomena such as confinement, screening, chiral
symmetries and $\cdots$. On a flat space--time, it is known that
in the massless Schwinger model via a peculiar two dimensional
Higgs phenomenon, the photon becomes massive and the Coulomb
force is replaced by a finite range force, giving rise to the
screening phase.  When the dynamical fermions are massive,
the model is not exactly soluble, but a semiclassical analysis reveals
a linear rise energy for opposites test charges $q$ and ${\bar q}$,
binding them into $q{\bar q}$ pairs \cite{col}.
By taking into accounts the finite distance corrections, one
can show that besides the linear confining term, the potential is
also composed of a screening term which is modified with respect to the
massless case \cite{ar}.

One of the interesting questions is how a curved background can modify
these effects? We think that this is an important question, because it
can be viewed as a first step in studying these physical effects in the
context of quantum gravity. Moreover, they may have applications in string
theory and quantum gravity coupled to nonconformal matter (note that the
kinetic term of the gauge fields spoils the conformal
invariance of the theory).

Although the Schwinger model on curved space--time has been
studied in several papers but the confining aspects of the model
is still unclear (for a discussion about the subtleties in
determining the confining phase of the Schwinger model on curved
space see \cite{gs}). For example the author of \cite{eb} has
suggested that the curvature of the space does not change the
confining behavior of the massive Schwinger model, and in
\cite{gh} it has been claimed that on a particular black hole, the
massless Schwinger model remains in screening phase. Also in
\cite{gs}, by comparison the role of temperature and the
curvature, it has been argued that the curvature may modify the
confining or screening nature of the Schwinger model. More
recently, this model has been investigated on a constant negative
curvature space--time (i.e., the Poincare half--plane) \cite{moh},
and it has been found that the confining feature of the Schwinger
model depends on the geometry of the space--time, for example, in
different regions of Poincare half--plane the system is in
different phases (i.e. confining or screening phases), which is in
contradiction with flat space--time results. Also in \cite{sad},
the confining aspects of the massless Schwinger model on de Sitter
space in different coordinates has been discussed and the results
have been extended to non--abelian situation.

In this paper we want to study the massive Schwinger model on a
general noncompact Riemann surface, and try to obtain, as much as
possible, information about the confinement in this background. We
consider only the effects of geometry on confinement phenomenon
and ignore nontrivial topologies.

In the Schwinger model the potential of external charges can be
obtained using several different methods, for example : (i)
Integrating over the fermionic fields (or equivalently over
bosonic fields in the bosonized version of the model), results an
effective action for the gauge fields, from which the potential
can be extracted by solving the corresponding equations of motion.
(ii) Integrating over gauge fields in the bosonized action and
then obtaining the static solutions of the equation of motion and
deducing the potential of the system as the difference between the
Hamiltonian in the presence and absence of external charges
respectively. In this paper we follow the second procedure.

The plan of the paper is as follows: In section 2, our main task
is to obtain the bosonic version of the massive Schwinger model on a
general non--compact curved background. To find this bosonic
representation, it is necessary to employ an appropriate normal
ordering description in defining the composite fields, which are
always present in the bosonic representations. As we will see,
this needs several change of field variables in order to reduce
our main model (interacting massive fermionic fields in curved
background) to a free massless fermionic field theory in flat
space--time. Then using the equivalence of this theory with a
massless bosonic field theory, we may obtain the appropriate
normal ordering description, from which our bosonic representation
can be found.

In section 3, by introducing two external charges $q$ and ${\bar
q}$ in the bosonized action, and by restricting ourselves to
static metrics (so that the static potential between probe charges
becomes meaningful), we obtain the energy of widely separated
external charges for small dynamical fermions mass.  Obviously,
the final result depends on the metric. For a specific example
(the full de Sitter space--time), we will show that one recovers a
confining behavior in the above mentioned limit. We also show that
when ${e'/q}\in Z$, where $e'$ ($q$) is the charge of external
(dynamical) charges, the external charges can not modify the
energy .This is the same result as the flat case. Note that in
section 2 where we use the path integral approach, it is more
convenient to consider the Euclidean signature, but in section 3, where
we are going to calculate the energy, it is better to work with Minkowskian
signature, in which the concept of energy is more natural.

To study the same problem but with external charges with finite
distance separation, in section 4 we restrict ourselves to the
case in which the gauge coupling is very large with respect to the
variation of the metric. The main reason for such a choice is
that, the massive scalar Green function, which is needed in
calculating the energy, is only known for a few number of
space--times. But if we restrict ourselves to the above mentioned
metrics, we can use the WKB approximation to find a general
expression for the energy of external charges. Like the flat case,
the energy takes a screening correction terms beside the confining
term. As a result we show that at $m=0$, the phase structure of
the system depends on the metric and in contrast to the flat case,
the model can be in confining phase in specific cases (such as the
full de Sitter space--time) which is in agreement with the result
of \cite{sad}.

\section{Bosonization of the massive Schwinger model}
All two dimensional spaces are conformally flat, hence any
noncompact Riemann surface is described by the metric
\be \l{1}
ds^2={\sqrt g}(dt^2+dx^2),
\ee
where $\sqrt g$ is the conformal
factor. On this space--time, the massive Schwinger model is
described by the partition function
\be \l{2}
Z=\int
DA_{\mu}D\psi D\psi^{\dagger}\exp\{-\int d^2x[{\sqrt g}
(\psi^{\dagger}\gamma^{\mu}(\partial_\mu -iqA_\mu)\psi +m
\psi^{\dagger}\psi )+{1\over 2\sqrt{g}}F^2]\},
\ee
where $\gamma^\mu \equiv {\hat \gamma}^ae^\mu_a$ are the curved
space counterparts of Hermitian Dirac gamma matrices: ${\hat
\gamma}^0=\sigma_2$ ; ${\hat \gamma}^1=\sigma_1$. $\sigma_i$ are
Pauli matrices. $\partial_{\mu}$ is the covariant derivative and
the zweibeins are defined through
\be \l{3}
g_{\mu
\nu}=e^a_\mu e^b_\nu \delta_{ab},\ \ \ \  g^{\mu \nu} =e^\mu
_ae^\nu _b\delta^{ab}.
\ee
The metric components are $g^{\mu
\nu}=(1/{\sqrt g}){\hat \delta}^{\mu \nu}$, where ${\hat
\delta}^{\mu \nu}={\hat \delta}_{\mu \nu}=diag(1,1)$. $q$ and $m$
are the charge and the mass of dynamical fermions respectively.
In eq.(\ref{2}), the dimension of
$q$ and $m$ is the inverse of the length dimension and $e$ is
dimensionless. The (dual) field strength $F$, is described through
$F={\hat \epsilon}^{\mu \nu}
\partial_{\mu}A_{\nu}$ in terms of gauge fields $A_{\mu}$,
where ${\hat \epsilon}^{\mu \nu}={\hat \epsilon}_{\mu \nu}$ and
${\hat \epsilon}^{01}=-{\hat \epsilon}_{10}=1$. The partition
function (\ref{2}) can be written as
\be \l{4}
Z=\sum_{k=0}^{\infty}{(-m)^k\o k!}<\prod_{j=1}^{k}\int d^2x_j
\sqrt{g({\bf {x_j}})} \psi^{\dagger}({\bf x}_j) \psi ({\bf x}_j)>,
\ee
where ${\bf x}=(x,t)$. In eq.(\ref{4}), the expectation values
are computed from the Lagrangian
\be \l{5}
L=\sqrt{g}\psi^{\dagger}\gamma^{\mu}(\partial_{\mu}-iqA_{\mu})\psi+
{1\o 2\sqrt{g}}F^2.
\ee
In terms of the new fermionic variables
${\tilde \psi}$ and ${\tilde \psi}^{\dagger}$:
$$
{\tilde \psi}=\exp(q\gamma_5 \varphi) \psi,
$$
\be \l{6}
{\tilde \psi}^{\dagger}= \psi^{\dagger}\exp(q\gamma_5 \varphi),
\ee
where $A_{\mu}={\hat \epsilon}_{\mu \nu}\partial_{\nu}\varphi$
and $\gamma_5=-i{\hat \gamma}^1{\hat \gamma}^0$, the Lagrangian (\ref{5})
can
be rewritten as \cite {na}
\be \l{7}
L=\sqrt{g}{\tilde \psi^{\dagger}}\gamma^{\mu}
\partial_{\mu}{\tilde \psi}+\sqrt{g}(-{\mu^2\over 2}\varphi \Delta \varphi
+{1\over 2}\varphi \Delta \Delta \varphi),
\ee
in which $\Delta $ is the Laplace--Beltrami operator: $\Delta=g
^{\mu \nu}\partial_{\mu}\partial_{\nu}$ and
\be \l{8}
\mu ={q\over \sqrt{\pi}}.
\ee
In this way the massless part has been became a free field theory on
a curved background with
an effective action
containing an anomalous term ($-(\mu^2/ 2)\sqrt{g}\varphi
\triangle \varphi$) coming from the Jacobian of the
transformation (\ref{6}) .
The term $(1/2)\sqrt{g} \varphi \Delta (\Delta -{\mu^2})\varphi$,
is the effective Lagrangian density of the gauge fields.
The fermionic part of the Lagrangian (\ref{7}) is free and hence is
invariant
under Weyl transformation: $g_{\mu \nu}\rightarrow \Omega^2 g_{\mu \nu}$
and
$\psi\rightarrow \Omega^{-{1\o 2}}\psi$. Choosing $\Omega =g^{-{1\o 4}}$,
one obtains
$$
g_{\mu \nu}\rightarrow {\hat \delta_{\mu \nu}},
$$
\be \l{9}
{\tilde \psi} \rightarrow \lambda =g^{1\o 8}{\tilde
\psi},\ \ \ \ {\tilde \psi}^{\dagger}\rightarrow
\lambda^{\dagger}= g^{1\o 8}{\tilde \psi}^{\dagger}.
\ee
Therefore
the partition function (\ref{2}) can be reduced to
\be \l{10}
Z=\sum_{k=0}^{\infty}{(-m)^k\o k!}<\prod_{j=1}^{k}\int d^2x_j
g^{1\o 4}({\bf x}_j)\lambda^{\dagger} ({\bf x}_j)\exp [-2q\gamma_5
\varphi ({\bf x}_j)]\lambda ({\bf x}_j)>,
\ee
where the expectation values are computed from
\be \l{11}
L=\lambda^{\dagger}{\hat \gamma}^a\partial_{a}\lambda +{1\o
2}\sqrt{g} \varphi\triangle(\triangle -\mu^2)\varphi.
\ee
Therefore the massless fermionic part of the theory is reduced to
a free fermionic field theory on a flat space--time. Using
$(\gamma_5)^{2n}=1$, it can be easily shown that
\be \l{12}
\lambda^{\dagger}\exp({-2q\gamma_5 \varphi})\lambda
=\lambda^{\dagger} [{1+\gamma_5 \o 2}\exp({-2q\varphi}) +{1-
\gamma_5 \o 2}\exp({2q\varphi})] \lambda.
\ee
The fermionic part
of (\ref{11}) is chirally invariant, hence the only non--zero
terms in  (\ref{10}) are those with equal number of $\sigma_{+}$
and $\sigma_{-}$ \cite{cl}, where
$\sigma_{\pm}=\lambda^{\dagger}(1\pm \gamma_5)\lambda /2$. Hence
\be \l{13}
Z=\sum_{k=0}^{\infty}{{(-m)}^{2k}\o
(k!)^2}<\prod_{j=1}^{k}\int d^2x_jd^2y_j g^{1\o 4}({\bf
x}_j)g^{1\o 4}({\bf y}_j)\exp[{2q\varphi ({\bf y}_j)}]
\exp[{-2q\varphi ({\bf x}_j)}] \sigma_{+}({\bf
x}_j)\sigma_{-}({\bf y}_j)>.
\ee
The expectation values of the
fermionic part can be established in the same manner as in the
flat case \cite{bo}:
\be \l{14}
\int D\lambda
D{\lambda^{\dagger}}\prod^{k}_{j=1}\sigma_{+}({\bf x}_{j})
\sigma_{-}({\bf y}_{j})\exp(-\int d^2x\lambda^{\dagger} {\hat
\gamma}^a{\partial_a}\lambda)=({1\o
2\pi})^{2k}{\prod^k_{i>j}({\bf x}_i- {\bf x}_j)^2({\bf y}_i-{\bf
y}_j)^2\o \prod_{i,j=1}^k({\bf x}_i- {\bf y}_j)^2}.
\ee
But the Green function of the massless scalar field $\phi({\bf x})$ is
$<\phi({\bf x})\phi({\bf y})>=
-1/(4\pi)\ln [{\tilde \epsilon}^2({\bf x}-{\bf y})^2]$,
where the mass ${\tilde \epsilon}^2 \ \ ({\tilde \epsilon}^2\rightarrow 0)$
is introduced to avoid infrared divergences. This is equivalent
to infrared renormalization of exponential of massless scalar field
discussed in \cite{sw} in operator language.
So using
$$
\int D\phi({\bf x})\exp[\Sigma_{j=1}^k{\alpha_j^2\o2}D({\bf x}_j,{\bf x}_j)]
\exp[\Sigma_{j=1}^k i\alpha_j\phi({\bf x}_j)]\exp[-{1\o 2}\int (\partial_{\mu}
\phi({\bf x}))^2d^2x] =
$$
\be
\delta_{\sum_{j=1}^k\alpha_j}\exp[-\sum_{i<j}^k\alpha_i \alpha_jD({\bf x}_i
,{\bf x}_j)],
\ee
in which we have used ${\tilde \epsilon}^2\rightarrow 0$ limit and
$D({\bf x}_i,{\bf x}_j)=-1/(4\pi)\ln({\bf x}_i-{\bf x}_j)^2$, one can write
the eq.(\ref{14}) in the form
$$
({1\o 2\pi})^{2k}\exp [2\pi \Sigma_{i=1}^{k}D({\bf x}_i,{\bf
x}_i)]\exp[2\pi \Sigma_{i=1}^{k}D({\bf y}_i,{\bf y}_i)]
\times
$$
\be \l{15}
\int D\phi({\bf x}) \exp\{2i\sqrt{\pi}[\sum^k_{j=1}
(\phi({\bf x}_j) -\phi({\bf y}_j))]\}\exp[-\int d^2x{1\o
2}(\partial_{\mu}\phi({\bf x}))^2].
\ee
Hence the eq.(\ref{13}) can be written as
\bea \l{16}
Z&=&\sum_{k=0}^{\infty}{m^{2k}\o (2\pi)^{2k}(k!)^2}
<\prod_{j=1}^{k}\int d^2x_jd^2y_jg^{1\o 4}({\bf x}_j)g^{1\o
4}({\bf y}_j) \exp[{2\pi D({\bf x}_j, {\bf x}_j)}]
\exp[{2\pi D({\bf y}_j,{\bf y}_j)}]\times \nonumber \\
& \;\;\;\;&
\exp[{-2q\varphi({\bf x}_j)}]\exp[{2q\varphi({\bf y}_j)}]\exp[2i\sqrt{\pi}
\phi ({\bf x}_j)]
\exp[-2i\sqrt{\pi} \phi ({\bf y}_j)]>,
\end{eqnarray}
where the expectation values is calculated from the following Lagrangian
(note that $\sqrt{g}g^{\mu \nu}={\hat \delta}^{\mu \nu}$):
\be \l{17}
L={1\o 2}\sqrt{g} \varphi \triangle (\triangle -\mu^2)\varphi +
{1\o 2}\sqrt{g} g^{\mu \nu} \partial_{\mu} \phi \partial_{\nu}\phi.
\ee

Now if we note that there is the charge conservation law in
contracting the composite fields $e^{i\beta \phi}$ \cite{cl}, then
the partition function (\ref{16}) can be written as
\bea \l {18}
Z&=&\sum^{\infty}_{k=0}{m^{2k}\o
(2\pi)^{2k}(k!)^2}<\prod_{j=1}^k\int d^2x_j
d^2y_jh({\bf x}_j)h({\bf y}_j)\times \nonumber \\
& \;\;\;\;& \exp[2i\sqrt{\pi}(\phi+{iq\o \sqrt{\pi}}
\varphi)({\bf
x}_j)]\exp[-2i\sqrt{\pi}(\phi +{iq\o \sqrt{\pi}}\varphi)({\bf y}_j)]>
\nonumber\\
&=& <\sum^{\infty}_{k=0}{1\o k!}\{{m\o \pi}\int d^2xh({\bf
x})\cos[2\sqrt{\pi}
(\phi({\bf x})+{iq\o \sqrt{\pi}}\varphi({\bf x})]\}^k>\nonumber \\
&=&<\exp\{{m\o \pi} \int d^2xh({\bf x})\cos[2\sqrt{\pi}(\phi({\bf
x})+{iq\o \sqrt{\pi}} \varphi({\bf x}))]\}>,
\eea
in which
\be \l{19}
h({\bf x})=g^{1\o 4}({\bf x})\exp[2\pi D({\bf x},
{\bf x})].
\ee
By changing the field variable $\phi\rightarrow
\phi-(iq/ \sqrt \pi)\varphi$, and using $A_{\mu}={\hat
\epsilon}_{\mu \nu}\partial_{\nu}\varphi$, the partition function
(\ref{18}) becomes (note that this change of field variable is
somehow inverse of (\ref{6}) and couples the matter and gauge
fields)
\begin{eqnarray} \l{20}
Z&=&\int DA_{\mu} D\phi \exp\{-\int d^2x\sqrt{g}
[{1\o 2}g^{\mu \nu}\partial_{\mu}\phi
\partial_{\nu} \phi
-i{q\over \sqrt{\pi}}\epsilon^{\mu \nu}A_{\mu}
\partial_{\nu}\phi
+{1\o 2g}F^2 \nonumber \\
& \;\;\;\;& - {m\o \pi}g^{-{1\o 4}}({\bf x})\exp[2\pi D({\bf x},{\bf x})]
{\rm cos}(2\sqrt{\pi}\phi({\bf x})]\},
\end{eqnarray}
where $\epsilon^{\mu \nu}=(1/ \sqrt{g}){\hat \epsilon}^{\mu \nu}$.
By integrating over the gauge fields one can see that the
effective Lagrangian of the field $\phi$ is
\be \l{21}
L_{{\rm
eff.}}={1\o 2}\sqrt{g} g^{\mu \nu}\partial_{\mu}\phi
\partial_{\nu}\phi
+{\mu^2 \o 2}\sqrt{g} \phi^2
-{m\o \pi}g^{1\o 4}\exp[2\pi D({\bf x},{\bf x})]\cos [2\sqrt{\pi}
\phi({\bf x})].
\ee
Now we want to convert the mass term in eq.({\ref{21}) to a more convenient
form, by absorbing covariantly the ultraviolet divergence in the mass $m$
or in operator language, by defining an appropriate normal ordering
prescription ($N_{\mu}$) with respect to the scale $\mu$ \cite{smil}.
To do so in the flat case, we write $\exp [2\pi D({\bf x},{\bf x})]$ as
\bea \l{22}
\exp[2\pi D({\bf x},{\bf x})]&=&\exp[-2\pi (G_{\mu}({\bf x},{\bf x})-
D({\bf x},{\bf x}))]\exp[2\pi G_{\mu}({\bf x},{\bf x})]
\nonumber \\
&=&\exp(\gamma +{\rm ln}{\mu \o 2})\exp[2\pi G_{\mu}({\bf x},{\bf x})],
\eea
where $G_{\mu}$ is the Green function of a massive scalar field with
mass $\mu$; which for $\sqrt{g}=1$ is $G_{\mu}^{\rm flat}=1/(2\pi)
K_{0}(\mu |{\bf x}-{\bf y}|)$. $K_0$ is the modified Bessel function
of the second kind and $\gamma$ is the Euler constant. We can now
absorb $\exp[2\pi G_{\mu}({\bf x},{\bf x})]$ in redefinition
(renormalization) of mass $m$:
\be \l{b}
m\psi^{\dagger}\psi=M{q\exp{\gamma}\o 2\pi^{3\over 2}}\cos(2\sqrt{\pi}\phi),
\ee
where $M\equiv m\exp[2\pi G_{\mu}^{\rm flat}({\bf x},{\bf x})]$. In operator
language \cite{cl}, this is equivalent to the well known result \cite{smil}:
\be \l{26}
m\psi^{\dagger}\psi
({\bf x})=-m{q \exp(\gamma)\o 2\pi^{3\o 2}}
N_{\mu}\cos[2\sqrt{\pi}\phi({\bf x})].
\ee
Now let us consider the
normal ordering in curved background. In this case, one can use a
covariant point--splitting method in order to regularize the
composite operator $\phi^2$ (our regularization method should
respect general covariance)
\be \l{27}
<\phi^2_{\rm {reg.}}({\bf x})>=
\lim_{{\bf x}\rightarrow {\bf x}'}[G_{\mu}({\bf x},{\bf x}')-
G_{\mu}^{DS}({\bf x},{\bf x}')],
\ee
where $G_{\mu}^{DS}({\bf
x},{\bf x}')$ (extracted from DeWitt--Schwinger expansion) is the
counterterm needed to regularize the ultraviolet divergence of the
Green function \cite{da}
\be \l{28}
G_{\mu}^{DS}({\bf x},{\bf
x}')=-{1\o 4\pi}[2\gamma + {\rm ln}(\mu^2 \epsilon^2)-{1\o
6}{R({\bf x})\o \mu^2} +O(\epsilon^2)].
\ee
$R$ is the scalar
curvature of the space and $\epsilon$ is one half of the proper
distance between ${\bf x}$ and ${\bf x}'$. ( Note that in the flat
case, $R=0$, eq.(\ref{28}) coincides with $\lim_{{\bf
x}'\rightarrow {\bf x}}G_{\mu}^{\rm flat} ({\bf x},{\bf x}')$, and
from (\ref{27}) we obtain the usual regularization which kills out
the loops.) Now if we follow the same steps as in the flat case,
but here using $G_{\mu}^{DS}({\bf x},{\bf x})$ instead of
$G_{\mu}^{\rm flat}({\bf x},{\bf x})$, we obtain
\be \l{30}
m\psi^{\dagger}\psi({\bf x})=-m{1\o \pi g^{1\o 4}({\bf
x})}\exp\{-2\pi [G_{\mu}^{DS}({\bf x},{\bf x})-D({\bf
x},{\bf x})]\} {\tilde N}_{\mu}\cos[2\sqrt{\pi} \phi({\bf x})],
\ee
in which we have defined the normal ordering ${\tilde
N}_{\mu}$ as
\be \l{31}
{\tilde  N}_{\mu}\cos[2\sqrt{\pi}\phi({\bf
x})]=\exp[2\pi G_{\mu}^{DS}({\bf x}, {\bf
x})]\cos[2\sqrt{\pi}\phi({\bf x})].
\ee

In the static curved space--time we can use a normal ordering
which like the flat case kills the loops (the vacuum
is now defined using the global time--like killing vector of the
static space--time \cite{fo}). In the same manner as the flat case,
we obtain
\be \l{33}
m\psi^{\dagger}\psi=-{m\o \pi g^{1\o
4}}\exp[2\pi (D- G_{\mu})({\bf x}, {\bf
x})]N_{\mu}\cos[2\sqrt{\pi} \phi({\bf x})].
\ee
It is easy to show that (\ref{33}) is equal to
(\ref{30}). From (\ref{33}) (or equivalently (\ref{30}) for static
space--time), we obtain
\be \l{34}
<\psi^{\dagger}\psi({\bf
x})>_{m=0}=-{1\o \pi g^{1\o 4}} \exp[-2\pi G({\bf x},{\bf x})],
\ee
where $G({\bf x},{\bf x})\equiv G_{\mu}({\bf x},{\bf x})-
D({\bf x},{\bf x})$. So, as we expect, the expectation
value $<\psi^{\dagger}\psi>$ is independent of the method of
regularization ( or normalization) of composite fields.

\section{ Quark-- antiquark potential in the massive Schwinger model
on static curved spaces: widely separated charges}

In this section we obtain the energy of two external static charges
introduced
into the massive Schwinger model. We consider an infinite static
conformally flat space-time, with trivial topology, described by the metric
\be \l{35}
ds^2=\sqrt{g(x)}(dt^2-dx^2).
\ee
The bosonized action of the massive Schwinger model,
in the presence of the covariantly conserved external current
\be \l{36}
J^0(x)={e'\o \sqrt{g}}(-\delta (x-b)+\delta (x-a)),\ \ \ \ J^1=0,
\ee
describing two opposite point charges $-e'$ and $e'$ located at $x=b$
and $x=a$ respectively, is
\bea \l{37}
S&=&\int d^2x[{1\o 2}\sqrt{g} g^{\mu \nu}\partial_{\mu}\phi \partial_{\nu}
\phi +{q\o \sqrt{\pi}}F\phi +{m\o \pi}g^{1\o 4}\exp [-2\pi G(x,x)]
{N}_{\mu}\cos(2\sqrt{\pi} \phi({\bf x})) \nonumber \\
& \;\;\;\; & +{1\o 2\sqrt{g}}F^2+{\tilde \eta}F],
\eea
where
\be \l{38}
{\tilde \eta}(x)=e'[\theta(x-a)-\theta(x-b)]=\cases {e', &$a<x<b$ \cr
0, &$x\notin [a,b].$ \cr}
\ee
Eq.(\ref{37}) is the Minkowskian version of (\ref{20}) in the
presence of external
current (\ref{36}). In deriving (\ref{37}) we have also used
eq.(\ref{33}).
Integrating over the field $F$ results
\bea \l{39}
S_{\rm {eff.}}&=&\int d^2x[{1\o 2}\sqrt{g}g^{\mu \nu}\partial_{\mu}\phi
\partial_{\nu}\phi+ {m \o \pi} g^{1\o 4}\exp [-2\pi G(x,x)]
{N}_{\mu}\cos(2\sqrt{\pi}
\phi({\bf x})) \nonumber \\
& \;\;\;\;&
-{q^2\sqrt{g} \o 2\pi }(\eta +\phi)^2],
\eea
in which we have defined $\eta =(\sqrt{\pi}/ q){\tilde \eta}$.
For widely separated charges, ${\tilde \eta}$ is equal to the
constant $e'$ in the whole space (in the next section
we will obtain the necessary corrections for finite charge separation distance).
If one changes the variable $\phi \rightarrow \phi -\eta$, finds
\bea \l{40}
S_{\rm {eff.}}&=&\int d^2x[{1\o 2}\sqrt{g}g^{\mu \nu}\partial_{\mu}\phi
\partial_{\nu}\phi + {m\o \pi}
g^{1\o 4}\exp [-2\pi G(x,x)]{N}_{\mu}\cos[2\sqrt{\pi}(\phi -\eta)({\bf x})]-
\nonumber \\
& \;\;\;\;&
{\mu^2\o 2}\sqrt{g}\phi^2].
\eea
For $(e'/ q)\in {\rm Z}$, the action is not modified by the
presence of external charges,
hence the energy is not changed. In other words, in this case, external
probes with
charge $e'$
are screened by dynamical fermions. The same effect
occurs in flat space--time \cite{col}.

Now let us restrict ourselves to  $|\phi|\ll 1$ scalar fields. In this
regime, the action (\ref{39})
becomes Gaussian and the classical solutions of the
action coincide with the quantum ones.
The classical equation of motion
for static field $\phi$ is
\be \l{41}
[\partial_1^2-\mu^2\sqrt{g} -
4\sqrt{\pi}m\Sigma \sqrt{g} \cos(2\sqrt{\pi}\eta)]\phi=-2
\sqrt{\pi}m\Sigma \sqrt{g} \sin(2\sqrt{\pi}\eta),
\ee
in which $\Sigma$ is defined through
\be \l{42}
\Sigma(x)\equiv \frac {1}{g^{1\o 4}\pi}
\exp[-2\pi G(x,x)].
\ee
The solution of eq.(\ref{41}) is
\be \l{43}
\phi=O^{-1}2\sqrt{\pi}m \Sigma \sqrt{g} \sin(2\sqrt{\pi}\eta),
\ee
where
\be
O=-\partial_1^2+\mu^2\sqrt{g}+4m\pi \Sigma \sqrt{g}\cos(2\sqrt{\pi}\eta).
\ee
Now as the eq.(\ref{43}) can be written as
\be \l{44}
\phi=[1-O^{-1}(O-
2\sqrt{\pi} m\Sigma\sqrt{g})]
\sin(2\sqrt{\pi}\eta),
\ee
therefore the condition $|\phi | \ll 1$ is satisfied when
$m\Sigma \ll \mu^2$.
The energy of the system is
\be \l{45}
E=\int T^0_0 dx=-\int L dx,
\ee
where $T^{\mu}_{\nu}$ is the energy momentum tensor and $E$ is the energy
measured by a static observer with respect to the coordinate (\ref{35}).
By substituting (\ref{43}) into (\ref{39}), we obtain
\be \l{46}
E(\eta)=-\int^b_a[2\pi m^2\sin^2(2\sqrt{\pi}\eta) \Sigma
\sqrt{g}O^{-1}\Sigma\sqrt{g}+
m\Sigma \sqrt{g}\cos(2\sqrt{\pi}\eta)]dx.
\ee
Up to the first order of $m$ and by considering $m\Sigma \ll \mu^2$,
one can arrive at
\be \l{47}
E(\eta)\cong -m\cos(2\sqrt{\pi}\eta)\int_{a}^{b}  \Sigma(x) \sqrt{g}\, dx.
\ee
Therefore the energy of external charges is
\bea \l{48}
E_{\rm ext.}&\equiv &
E(\eta)-E(0)\nonumber \\
&=&m[1-\cos(2\pi{e'\o q})]\int^b_a \Sigma(x) \sqrt{g} dx.
\eea
This completes our general result for the energy of widely separated external
charges  $\pm e'$ in a static curved background.
This result coincides exactly with the result of \cite{sad}, in which
the same problem has been discussed using the
fermionic action.
As is clear from (\ref{47}), the
determination of the energy is now complicated by the presence
of the conformal factor ($\sqrt{g}$) and $\Sigma$, which the latter can be
expressed in terms of Seeley
DeWitt coefficients although not explicitly known for general curved
space--times.
To obtain an insight about this result, let us
consider a specific example.
\vskip 0.5cm
{\bf {Example: complete de Sitter space--time}}
\vskip 0.5cm
Consider the following geodesically complete
space--time \cite{mi}
\be \l{49}
ds^2=(1+{x^2\o \lambda^2})dt^2-{dx^2\o {1+{x^2/\lambda^2}}},
\ee
which represents the full two--dimensional de Sitter space--time,
with scalar curvature $R={2/ \lambda^2}$.
By the change of coordinate $x\rightarrow x(r)$, where
$dr^2={dx^2/ (1+{x^2/\lambda^2})^2}$ or
$r=\lambda \cos^{-1}[\lambda /{(x^2+\lambda^2)}^{1\o 2}]$, this space
takes the conformally flat form
\bea  \l{50}
ds^2&=&(1+{x^2\o \lambda^2})(dt^2-dr^2)\nonumber \\
&=&{1\o \cos^2({r \o \lambda})}(dt^2- dr^2) ,\ \ \ \,
-{\pi\o 2}< {r\o \lambda}<{\pi\o 2}.
\eea
On spaces with constant positive
curvature $R$, $G_{\mu}({\bf x},{\bf y})$ is \cite{rt}
\be \l{51}
\lim_{{\bf x}\rightarrow {\bf y}} G_{\mu}({\bf x},{\bf y})=
-{1\o 4\pi}\{2\gamma +{\rm ln}({R\epsilon^2
\o 2}) +\Psi({1\o 2}+\alpha)+\Psi({1\o 2} -\alpha)\},
\ee
where $\alpha^2={1/ 4}+{2\mu^2}/R$ and $\Psi$ is the digamma function.
As a consequence  $\Sigma$ is
\be \l{52}
\Sigma={1\o \pi}\exp[\gamma +{1\o 2}{\rm ln}({R\o 2}) +
{1\o 2}\Psi({1\o 2}+\alpha)
+{1\o 2}\Psi({1\o 2} -\alpha)].
\ee
The energy of widely separated external charges is then
\be \l{53}
E_{\rm ext.}= m[1-\cos(2\pi {e'\o q})]\Sigma (b-a).
\ee
$b$ and $a$ are expressed in terms of $x$ coordinate.
Although the energy is not linear in terms of charge separation distance
$d= \int^b_a dx/ \sqrt{1+x^2/ \lambda^2}$, but for
$d\rightarrow \infty$
we have
$E\rightarrow \infty$, hence the system is in confining phase.

\section{\bf {Energy of finite separated charges}}

If we consider
(\ref{47}) for $m=0$, it gives $E_{\rm ext.}=0$, which is not
right as it does not contain the finite separation contributions.
In this section we want to study the case in which the external charges
have finite separation, although the energy (\ref{47}), is yet the dominant
long range term.
In this case, the function ${\tilde \eta}$
can not be considered as a constant in the whole of the space.
As has been mentioned in section one, in flat space--time the energy
expression, beside the confining term, consists also of correction terms
which are due to the screening behavior of the system. Here we want
to calculate
these correction terms for a static curved space--time.

Let us consider
the general expression (\ref{39}), in which we must now treat ${\tilde \eta}$
as a $x$-dependent expression. Before any calculation, we must say something
about $G({\bf x},{\bf x})$ in (\ref{39}). As was pointed earlier, we don't
know the explicit form of this function for a general curved space--time.
So we will restrict ourselves to a specific condition for the metric
(or the gauge coupling). We first note that on the conformally
flat space--time
(\ref{1}), $G_{\mu}({\bf x},{\bf x}')$ satisfies
\be \l{54}
({d^2\o dx^2}+{d^2\o dt^2}-\mu^2\sqrt{g})G_{\mu}({\bf x},{\bf x}')=
-\delta({\bf x},{\bf x}').
\ee
Using the expansion
\be \l{55}
G_{\mu}({\bf x},{\bf x}')={1\o 2\pi}\int G_{\mu}(k,x,x')e^{ik(t-t')}dk,
\ee
we obtain the following equation for $G_{\mu}(k,{\bf x},{\bf x}')$
\be \l{56}
[{d^2\o dx^2}-(k^2+\mu^2\sqrt{g})]G_{\mu}(k,x.x')=-\delta(x,x').
\ee
Now if we consider the large gauge coupling regime, such that
\be  \l{57}
{dg^{1\o 4}\o dx}\ll \mu \sqrt{g},
\ee
then the solution of (\ref{56}) ( at zeroth order of WKB approximation) is
\be \l{58}
G_{\mu}(k,x,x')={1\o 2(k^2+\mu^2\sqrt{g(x)})^{1\o 4}
(k^2+\mu^2\sqrt{g(x')})^{1\o 4}}\exp[-|\int^{x'}_x
\sqrt{k^2+\mu^2\sqrt{g(u)}}
du|].
\ee
In the limit ${\bf x}\rightarrow {\bf x'}$, $G_{\mu}({\bf x},{\bf x'})$ is
found to be
\be \l{59}
\lim_{{\bf x}\rightarrow {\bf x}'}G_{\mu}({\bf x},{\bf x}')={1\o 2\pi}
\lim_{(x,t)\rightarrow (x',t')}K_0\{\mu^2\sqrt{g}[
(x-x')^2+(t-t')^2]\}^{1\o 2}.
\ee
where $\delta s=g^{1\o 4}(x)[(x-x')^2+(t-t')^2]^{1\o 2}$ is the
distance between two points ${\bf x}$ and ${\bf x}'$ when
${\bf x}'\rightarrow {\bf x}$.
Now using $D({\bf x},{\bf x}')=-(1/ 2\pi){\rm ln}(\delta s)$
and by considering the behavior of $K_0$ for small arguments,
$G({\bf x},{\bf x})$ (introduced after eq.(\ref{34})) is found to be
\be \l{60}
2\pi G({\bf x},{\bf x})=-\{\gamma+{\rm ln}{\mu g^{1\o 4}\o 2}\}.
\ee
Thus eq.(\ref{34}) yields
\be  \l{61}
<\psi^{\dagger}\psi>_{m=0}=-{q\exp(\gamma)\o 2\pi^{3\o 2}}.
\ee
Note that in this approximation, the metric factor in eq.(\ref{34}) is canceled
out by the corresponding term in $G({\bf x}, {\bf x})$, and the final result
is the same as the flat case as one expects from the results of \cite{wi}.

Now let us go back to eq.(\ref{39}). If we restrict ourselves to the regime
(\ref{57}),
in which eq.(\ref{60}) has been obtained, and change the variable $\phi$ as
$\phi \rightarrow \phi -\eta $, the effective Lagrangian (\ref{39}),
for static
fields with $|\phi -\eta |\ll 1$ condition, is reduced to the following
Lagrangian
\bea \l{62}
L&=&{1\o 2}\phi(\partial_1^2-\mu^2 \sqrt{g} -4\pi m\Sigma \sqrt{g})
\phi -\phi(\partial_1^2-4\pi m\Sigma \sqrt{g} )\eta
\nonumber \\
& \;\;\;\;&
+{1\o 2}\eta \partial_1^2\eta -2\pi m\Sigma\sqrt{g} \eta^2.
\eea
Using (\ref{42}) and (\ref{60}), $\Sigma$ is obtained as
$\Sigma= q\exp(\gamma)/ (2\pi^{3/ 2})$.
The field equation is
\be \l{63}
\phi= {-\partial_1^2+4\pi m\Sigma \sqrt{g}\o -\partial_1^2+\mu^2 \sqrt{g}
+4\pi m\Sigma\sqrt{g}}\eta.
\ee
Using the following identity which holds at large coupling limit
\bea \l{64}
&\int^x&{[\sqrt{g(x)}]^\alpha \o [\mu^2\sqrt{g(x)}+4\pi m\Sigma
\sqrt{g(x)}]^\beta}
\exp[-\int^x_a\sqrt{\mu^2\sqrt{g(u)}+4\pi m\Sigma \sqrt{g(u)}}du] dx
\nonumber \\
& \;\;\;\;&  = -{[\sqrt{g(x)}]^\alpha \o [\mu^2 \sqrt{g(x)} +4 \pi m\Sigma
\sqrt{g(x)}]^{\beta +{1\o 2}}}\exp [-\int^x_a \sqrt{\mu^2 \sqrt{g(x)} +4\pi
 m\Sigma \sqrt{g(x)}} \, dx]+c,
\eea
in which $c$ is an arbitrary constant, we obtain
\be \l{65}
\phi (x)=\cases{ \sqrt{\pi}{e'\o q}{\mu^2\o 2}[F(x,b)-F(x,a)], &$x>b$, \cr
\sqrt{\pi}{e'\o q}- \sqrt{\pi}{e'\o q}{\mu^2\o 2}[2F(x,x)-
F(x,a)-F(x,b)], &$a<x<b $, \cr
-\sqrt{\pi}{e'\o q}{\mu^2\o 2}[F(x,b)-F(x,a)], &$x<a$, \cr }
\ee
where
\be \l{66}
F(x,y)={\sqrt{g(y)}\o f^{1\o 4}(x)f^{3\o 4}(y)}\exp[-|\int^y_xf^{1\o 2}(u)du|],
\ee
and
\be \l{67}
f(x)=(\mu^2+4\pi m\Sigma)\sqrt{g(x)}.
\ee
In deriving (\ref{65}), the Green function (\ref{58}) has been used.
For a fix $x$, $F(x,y)$ is an increasing function
of $y$
in $y<x$ region and decreasing in $y>x$. So $F(x,y)\leq F(x,x)$.
But $\mu^2F(x,x)=\mu^2/ (\mu^2+4\pi m\Sigma)<1$, which is not
necessarily small. Therefore
the condition $|\phi -\eta|\ll 1$ is satisfied only when the factor $e'/q$
is very small. The same arguments holds when $e'/q$ is close to an integer
number.

Putting (\ref{65}) back into the Lagrangian (\ref{62}) gives
\be \l{68}
E_{\rm ext.}={\mu^2 \o 2}\{\int^b_a {[\sqrt{g}\eta^2-\mu^2\sqrt{g}\eta
{1\o -\partial^2_1 +\mu^2\sqrt{g}+4\pi m\Sigma\sqrt{g}}\sqrt{g}\eta]}dx\}.
\ee
Using  (\ref{64}) we arrive at
\bea \l{69}
E_{\rm ext.}&=&\pi ({e'\o q})^2\{(1-{\mu^2 \o \mu^2 +4\pi m\Sigma})
{\mu^2 \o 2}\int_{a}^{b}\sqrt{g(x)} \, dx
\nonumber \\
& \;\;\;\;& +{\mu^4\o 4}[{g(a)\o f^{3\o 2}(a)}+{g(b)\o f^{3\o
2}(b)}-{2\sqrt{g(a)g(b)} \o (f(a)f(b))^{3\o
4}}\exp(-\int^b_af^{1\o 2}(u)du)]\} . \eea
The energy of external
charges is composed of two parts: The first part is proportional
to $\int_{a}^{b} \Sigma \sqrt{g(x)} \, dx$ and when $m\Sigma\ll
\mu^2$, coincides with the confining term (\ref{47}) (in the limit
$e'\ll q$). On the flat space--time the remaining terms are due to the
screening of external charges by dynamical
fermions.
On the curved space, the problem is more
complicated and we can have a confining situation even in the
massless Schwinger model. To see this, assume that $m=0$. The
energy of external charges is then
\be \l{70}
E_{\rm
ext.}={e'^2\o 4\mu}[g^{1\o 4}(a)+g^{1\o 4}(b)-2g^{1\o 8}(a)
g^{1\o 8}(b) \exp(-\mu \int_{a}^{b} g^{1\o 4}(u) \, du)],
\ee
which for largely separated external charges, in contrast to the
flat case, does not tend to a constant, and depends on the value
of the metric at the position of external charges. The  condition
of validity of the eq.(\ref{70}) is only (\ref{57}). In other
words, in $m=0$ there is no need to condition $(e'/q)\ll 1$ (or
$|\phi -\eta|\ll 1$) in deriving (\ref{70}). For example in the
massless Schwinger model on de Sitter space (\ref{50}), the energy
of largely separated charges (in the coordinate $x$) is
\be \l{71}
E_{\rm ext.}={e'^2\o 4\mu}[\sqrt{1+{a^2\o \lambda^2}}
+\sqrt{1+{b^2\o \lambda^2}}].
\ee
Although this energy is not
linear in terms of charge separation distance, but for
$d\rightarrow \infty$, $E$ is infinite and the system is in
confining phase.

From eq.(\ref{70}), we can conclude that in contrast to the flat case, in
which the energy of external charges is only a linear function
of charge separation
distance, on the curved space--time the energy depends also on the position
of external charges. As it is clear, the infinity of the energy in a
confining situation is related to the separation of charges, but the
increasing rate of the energy with distance is not unique in all regions.
For example in the region $x\simeq x_{\rm sing}$, where
$x_{\rm sing.}\in R\bigcup {\pm \infty}$ is a point at which the metric
is singular, two external charges located close together can have
a finite energy but by moving one of the charges, the energy increases
very rapidly.

\vskip 1cm

\noindent{\bf Acknowledgement}\\
M. Alimohammadi would like to thank
the research council of the
University of Tehran for partial financial support.

\end{document}